\documentclass[superscriptaddress,twocolumn,showpacs,preprintnumbers,prl]{revtex4}
\usepackage{times,xspace}
\usepackage{amsbsy,amssymb,amsmath,bm}
\usepackage{graphicx,color,epsfig}
\usepackage{fancyhdr}

\def\bbbc{{\mathchoice {\setbox0=\hbox{$\displaystyle\rm C$}\hbox{\hbox
to0pt{\kern0.4\wd0\vrule height0.9\ht0\hss}\box0}}
{\setbox0=\hbox{$\textstyle\rm C$}\hbox{\hbox
to0pt{\kern0.4\wd0\vrule height0.9\ht0\hss}\box0}}
{\setbox0=\hbox{$\scriptstyle\rm C$}\hbox{\hbox
to0pt{\kern0.4\wd0\vrule height0.9\ht0\hss}\box0}}
{\setbox0=\hbox{$\scriptscriptstyle\rm C$}\hbox{\hbox
to0pt{\kern0.4\wd0\vrule height0.9\ht0\hss}\box0}}}}

\newcommand{\ignore}[1]{}
\newcommand{\mComment}[1]{}
\newcommand{\gComment}[1]{}
\newcommand{\jComment}[1]{}
\newcommand{\rComment}[1]{}
\newcommand{\lComment}[1]{}
\renewcommand{\gComment}[1]{\textcolor{magenta}{Gerardo: #1}}
\begin{document}
\title{The Canted Spiral: An Exact Ground State of XXZ Zigzag Ladders}
\author{C. D. Batista} 
\affiliation{Theoretical Division, Los Alamos National Laboratory,
Los Alamos, New Mexico 87545}
\date{\today}

\begin{abstract}
We derive the exact ground states for a one dimensional family of $S=1/2$ XXZ 
Hamiltonians on the zigzag ladder. These states exhibit true long 
range spiral order that spontaneously breaks the U(1) invariance of the Hamiltonian. 
Besides breaking a continuous symmetry in $d=1$, this spiral ordering has a ferromagnetic 
component along the symmetry axis that can take any value between zero and full saturation. 
In this sense, our canted spiral solutions are a 
generalization of the SU(2) Heisenberg ferromagnet to non-zero ordering wave-vectors of 
the transverse spin components. We extend this result to the $d=2$ anisotropic triangular lattice. 
\end{abstract}

\maketitle

\begin{figure}[thb]
\vspace*{-0.8cm}
\hspace*{-1.0cm}
\includegraphics[angle=0,width=10cm]{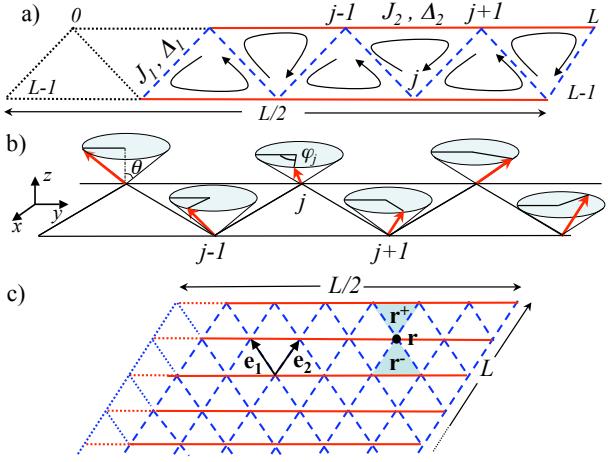}
\vspace{-1.5cm}
\caption{a) Zigzag ladder. The dotted lines indicate that we are using PBC's. 
The arrows show the circulation of the spin
and the  electrical orbital currents that result from the nonzero vector
and scalar chiralities of the canted spiral solution. b) Canted spiral solution.
The arrow at site $j$ corresponds to $\langle {\tilde {\Psi}}_{\theta,\varphi_0}|{\bf S}_j| {\tilde {\Psi}}_{\theta,\varphi_0} \rangle$.
c) Extension of the zigzag ladder to $d=2$. ${\bf e}_1$ and ${\bf e}_2$ are primitive vectors.${\bf r}_+$ and ${\bf r}_-$ 
denote the triangular plaquettes that are above and below the site ${\bf r}$.}
\label{zz}
\end{figure}

The search for chiral phases in one dimensional frustrated magnets has been very active 
during the last ten years \cite{Nersesyan98,Kolezhuk00,Hikihara01,Kolezhuk05,Okunishi08,McCulloch08}. 
This interest was triggered by the prediction of a ground state with non-zero vector spin chirality, 
$\langle {\bf S}_{j} \times {\bf S}_{j+1} \rangle \neq 0$, for the $J_1$, $J_2$  XXZ chain
with $|J_1| \ll J_2$ ($J_1$ and $J_2$ are the nearest and next-nearest exchange interactions). 
The $J_1$, $J_2$ chain is equivalent to a zigzag ladder (see Fig.\ref{zz}a) and it has also attracted a lot
of interest during the last decades  \cite{White96,Allen97}. One of the main reasons for the continuiung 
fascination generated by this model is in the interplay between geometric frustration and strong quantum fluctuations
that leads to a rich and exotic variety of physical phenomena. 

The vector chirality has to be distinguished from the scalar chirality, 
$\langle{\bf S}_{j-1} \cdot {\bf S}_{j} \times {\bf S}_{j+1} \rangle$, which breaks different discrete symmetries and
is associated with different physical quantities \cite{Lev08}. 
As it is mentioned in Ref.{\cite{Kolezhuk05}}, classical states with spontaneously broken chirality only exist together 
with helical long range order.
While the vector chirality distinguishes left and right spirals, the sign of the scalar chirality 
distinguishes between positive and negative canting angles for a given spiral orientation. The helical
order breaks the continuous symmetry of global spin rotations along the $z$-axis. Consequently, the existence 
of long range helical order is in most cases precluded by zero point fluctuations of $d=1$ quantum systems 
(Mermin-Wagner theorem \cite{Mermin66}). On the other hand, chiral orderings are allowed because 
they only break discrete symmetries. It is for this reason that chiral orders in quantum spin systems can be thought as remnants of the 
helical order in classical systems. This has been one of the main motivations for finding chiral orders in
quantum spin Hamiltonians whose ground state exhibits helical order in the $S \to \infty$ limit.

The zigzag XXZ ladder is one of the simplest spin models whose classical counterpart has a helical ground state 
in a certain region of exchange parameters. The original proposal of a chiral ground state for the quantum version of this model was based on a 
mean field treatment of the bosonized Hamiltonian \cite{Nersesyan98}. A similar approach was used by Kolezhuk and Vekua \cite{Kolezhuk05} 
to obtain a field induced chiral state in the  Heisenberg (XXX) zigzag ladder. These mean field approaches were later
validated by numerical simulations \cite{Okunishi08,McCulloch08}, but there are some regions of the quantum phase diagram
were the situation is still unclear \cite{McCulloch08}. In general, the numerical detection of these phases is very challenging due 
to the lack of commensuration between the dominant wave-vector of the spin-spin correlator and the finite size chains that can be solved 
numerically \cite{Aligia00}.  It is therefore desirable to find exact analytical solutions for the chiral ground states and compute the 
most relevant correlation functions without making any approximation.

In this Letter we consider a one dimensional family of S=1/2 XXZ zigzag ladders for which we derive an exact ground state 
subspace with a very unusual property: {\it it is the semi-classical version of a canted spiral phase} (see Fig.\ref{zz}b). 
In particular, this implies that 
the ground state has true long range helical order, i.e., it breaks spontaneously the continuous U(1) symmetry of global spin
rotations along the $z$-axis.
Moreover, the ground state is still a canted spiral on finite size ladders for a discrete set of ratios, $J_1/J_2$, between the two exchange
constants. 
The most curious of aspect  of these solutions is that in contrast to the usual example of the SU(2) Heisenberg ferromagnet (the SU(2) symmetry is also
spontaneously broken at $T=0$), the order parameter of the canted spiral {\it does not commute} with the Hamiltonian. We will see
that this observation is related to the emergence of an  SU(2) symmetry in the ground state subspace \cite{Batista04}. 
The nonzero scalar and vector chiralites are an epiphenomenon of the helical solution. They are also computed in an exact
way by exploiting the semi-classical or unentangled nature of the solution. We also provide an exact calculation of
a branch of gapless single-magnon excitations with a quadratic low-energy dispersion around the spiral wave-vectors $\pm Q$.
Finally, we extend our exact canted spiral solution to the case of an XXZ Hamiltonian defined on a
two dimensional version of the zigzag ladder (see Fig.\ref{zz}c). This is indeed the magnetic lattice of the frustrated quantum magnet
Cs$_2$CuCl$_4$ \cite{Coldea01}.

We start by considering an XXZ model on the zigzag
ladder with $L$ sites ($0\leq j \leq L-1$) depicted in Fig.\ref{zz},
\begin{eqnarray}
H = \sum_{j,\nu} J^z_{\nu}  (S^z_{j+\nu} S^z_j - \frac{1}{4})+ 
\frac{J_{\nu}}{2}( S^+_{j+\nu}S^-_{j} + S^-_{j+\nu}S^+_{j}),
\label{Ham}
\end{eqnarray}
where $\nu=1,2$, $0 \leq j \leq L-1 $, $L \equiv 0$ (periodic boundary conditions), and $S^{\pm}_j=S^x_j \pm i S^y_j$. We 
will assume that the ratio of exchange constants satisfies the equation
\begin{equation}
\cos{Q} = - \frac{J_1}{4 J_2}
\label{cos}
\end{equation}
where $Q= 2 \pi n /L$ ($0 \leq n \leq L-1$) is a wave-vector in the Brillouin zone (BZ) of the ladder. 
We will also assume $J_2 >0$, so Eq.(\ref{cos}) defines a set of $L/2+1$ different ratios $J_1/J_2$ for
even $L$ (the ratio is the same for $\pm Q$). In the thermodynamic limit, $L\to \infty$, $Q$ can take 
any arbitrary value between $0$ and $\pi$ and  Eq.(\ref{cos}) always has a solution for $|J_1| \leq 4 J_2$. 

We are particularly interested in the one dimensional family of Hamiltonians 
defined by $J^z_{\nu} = \Delta_{\nu} J_{\nu}$ with 
\begin{eqnarray}
\Delta_1 &=& \cos{Q} = - \frac{J_1}{4 J_2},
\nonumber \\
\Delta_2 &=&\cos{2Q} = - 1 + \frac{J^2_1}{8 J^2_2}.
\label{dd}
\end{eqnarray}
{\it Ground States}.The next step is to provide a subspace of exact ground states for this Hamiltonian family. 
For this purpose we will define the vacuum, $| \emptyset \rangle= \bigotimes_j |\! \! \downarrow \rangle_j$, as the fully polarized 
state with all the spins 
down: $S^z_j | \emptyset \rangle = -(1/2) | \emptyset \rangle \;\; \forall \;\; 0 \leq j \leq L-1$.
In addition, we introduce the spin operators in momentum space:
\begin{eqnarray}
{\cal S}^{z}_q = \frac{1}{\sqrt{L}}  \sum_j e^{iqj} S^{z}_j, 
\;\;\;\;
{\cal S}^{+}_q = \frac{1}{\sqrt{L}}  \sum_j e^{iqj} S^{+}_j, 
\label{four}
\end{eqnarray}
and ${\cal S}^{-}_q= ({\cal S}^{+}_q)^{\dagger}$.
These operators obey the following commutation relations:
\begin{equation}
[{\cal S}^+_k,{\cal S}^-_q] = \frac{2}{\sqrt{L}} {\cal S}^z_{k-q},
\;\;\;\;\;
[{\cal S}^z_q,{\cal S}^+_k] = \frac{1}{\sqrt{L}} {\cal S}^+_{k+q}.
\label{cr}
\end{equation}
We claim that the set of linearly independent states 
\begin{equation}
|\Psi_p \rangle = \sqrt{ \frac{L^{p} (L-p)!}{ p!  L!}} ({\cal S}^{+}_Q)^p  | \emptyset \rangle
\label{gst}
\end{equation} 
are exact ground states of $H$ for $0\leq p \leq L$. The normalization prefactor, $\langle \Psi_p | \Psi_p \rangle=1$,
is obtained by noticing that  $L^{p/2}({\cal S}^{+}_Q)^p  | \emptyset \rangle$ is a linear combination of 
$\binom{L}{p}$ states with coefficients that have the same absolute value equal to $p!$.
 
There are different ways of demonstrating our claim. Here we will use an algebraic procedure
that unveils an emergent SU(2) symmetry of $H$. In first place, we will demonstrate that 
the states $|\Psi_p \rangle$ are degenerate eigenstates of $H$. For this purpose it is 
convenient to express $H$ in momentum space:
\begin{equation}
H = \sum_{q} (J^z_q {\cal S}^z_q {\cal S}^z_{-q} + J_q {\cal S}^{+}_q {\cal S}^{-}_q),
\end{equation}
where 
\begin{eqnarray}
J^z_q &=& J_1 \Delta_1 \cos{q} + J_2 \Delta_2 \cos{2q},
\nonumber \\
J_q &=& J_1 \cos{q} + J_2 \cos{2q}.
\end{eqnarray}
From this expression we derive the following commutation relation by using Eqs.(\ref{dd}):
\begin{equation}
[S^{+}_{Q},H]= \frac{i}{\sqrt{L}} \sum_{l=1,L} e^{ilQ} S^{+}_l a_l,
\label{comm}
\end{equation}
with
\begin{equation}
a_l = \sum_{\nu=1,2} J_\nu \sin{\nu Q} \; (S^z_{l+\nu}- S^z_{l-\nu}).
\end{equation}
From Eq.(\ref{Ham}), it is evident that 
\begin{equation}
H | \emptyset \rangle =0.
\label{em}
\end{equation}
In addition, our definition of the vacuum state, 
$S^z_j | \emptyset \rangle = -(1/2) | \emptyset \rangle \;\; \forall \;\; 0 \leq j \leq L-1$,
implies that $(S^z_j-S^z_l)| \emptyset \rangle=0 \;\; \forall \;\; 0 \leq j,l \leq L-1$, 
and $a_l | \emptyset \rangle=0$.
Consequently,
\begin{equation}
[S^{+}_{Q},H] | \emptyset \rangle =0.
\label{f1}
\end{equation}
Finally, by using Eqs.(\ref{four}) and (\ref{comm}) it is easy to verify that 
\begin{equation}
[S^{+}_{Q},[S^{+}_{Q},H]] | \emptyset \rangle =0.
\label{f2}
\end{equation}
The combination of Eqs.(\ref{f1}) and (\ref{f2}) leads to a very important result:
\begin{equation}
[S^{+}_{Q},H] (S^{+}_{Q})^p| \emptyset \rangle =0.
\label{f11}
\end{equation}
This result implies that $S^{+}_{Q}$ is an infinitesimal generator of a continuous symmetry of $H$
when $H$ is restricted to the subspace ${\cal G}_Q$ generated by the states $|\Psi_p \rangle$ with $0 \leq p \leq L$:
\begin{equation}
[S^{+}_{Q},P_{{\cal G}_Q} H P_{{\cal G}_Q} ] =0,
\label{su2}
\end{equation} 
where $P_{{\cal G}_Q}$ is the projector on the subspace ${\cal G}_Q$. Moreover, Eqs.(\ref{em}) and (\ref{f11})
imply that $H (S^{+}_{Q})^p| \emptyset \rangle =0$ or 
\begin{equation}
H |\Psi_p \rangle =0 \;\; \forall \;\; 0 \leq p \leq L.
\end{equation}
This concludes our first step. The second step is to demonstrate that the eigenstates $|\Psi_p \rangle$ are 
{\it ground states} of $H$. Since the corresponding eigenvalues are equal to zero, we just need to demonstrate that
$H$ is semi-positive definite. This can be done by rewriting the Hamiltonian in the following way:
\begin{equation}
H = \sum_j H_j,
\label{sh}
\end{equation}  
with $H_j = h_{2}(j-1,j+1) + h_{1}(j-1,j) + h_{1}(j,j+1)$
\begin{eqnarray}
\!\!\!\!\!\! h_2(l,n) &=& J^z_2  (S^z_{l}S^z_{n} - \frac{1}{4})+ 
\frac{J_2}{2}( S^+_{l}S^-_{n}+ S^-_{l}S^+_{n}) 
\nonumber \\
\!\!\!\!\!\! h_1(l,n)&=& \frac{J^z_1}{2}  (S^z_{l}S^z_{n} - \frac{1}{4})+ 
\frac{J_1}{4}( S^+_{l}S^-_{n}+ S^-_{l}S^+_{n}) 
\end{eqnarray} 
For $\Delta_1$ and $\Delta_2$ given by Eqs.(\ref{dd}), we obtain that 
\begin{equation}
H_j = (J_2+J_1^2/8J_2) P_j
\label{klein}
\end{equation}
where $P_j$ is a projector on the two-dimensional subspace ${\cal E}(j-1,j,j+1)$ generated by the states:
\begin{eqnarray}
| \xi_{j \uparrow}\rangle &=&  \gamma (2 \cos{Q}  S^{+}_{j-1} S^{+}_{j+1}
- S^{+}_{j} S^{+}_{j+1}- S^{+}_{j-1} S^{+}_{j}) |\emptyset\rangle_j,
\nonumber \\
| \xi_{j \downarrow}\rangle &=&  \gamma (2 \cos{Q}  S^{+}_{j} 
- S^{+}_{j-1} -  S^{+}_{j+1}) |\emptyset\rangle_j,
\label{subs}
\end{eqnarray}
with $\gamma= 1/\sqrt{4+2 \cos{2Q}}$ and 
$|\emptyset \rangle_j = |\! \! \downarrow \rangle_{j-1} \otimes |\! \! \downarrow \rangle_j
 \otimes |\!\!\downarrow ~ \!\!\rangle_{j+1}$.
Given that $J_2>0$, the combination of Eqs.(\ref{sh}) and (\ref{klein}) implies that 
$H$ is semi-positive definite.
This concludes the demonstration of our main claim: {\it the states $|\Psi_p \rangle$ generate a ground
state subspace ${\cal G}_Q$ of $H$}. For $Q \neq 0, \pi$, there are two ground state subspaces, 
${\cal G}_Q$ and ${\cal G}_{\bar Q}$ (${\bar Q}=-Q$), that correspond to right and left  
spiral solutions as we will see below. In addition to these solutions, there are other linearly
independent ground states in the ${\cal S}^z=0$ subspace that will be presented elsewhere \cite{Batista10}.

{\it Order Parameter.} The exact ground subspace ${\cal G}_Q$  contains a spiral order parameter that breaks spontaneously a continuous 
symmetry of $H$: this is the U(1) symmetry of global spin rotations around the $z$-axis. To see this we just need
to choose an appropriate linear combination of the $|\Psi_p \rangle$ ground states:
\begin{equation}
|{\tilde {\Psi}}_{\theta,\varphi_0} \rangle = 
e^{\frac{ - \theta \sqrt{L}}{2} (e^{i\varphi_0} {\cal S}^+_Q - e^{-i\varphi_0}{\cal S}^-_Q )} | \Psi_L \rangle.
\end{equation}
To verify that $|{\tilde {\Psi}}_{\theta,\varphi_0} \rangle \in {\cal G}_Q$ we just need to 
notice that ${\cal G}_Q$ is an invariant subspace of ${\cal S}^+_Q$ and ${\cal S}^-_Q$, something that 
will become evident when we discuss the emergent SU(2) symmetry of $H$. It is easy to verify that 
$|{\tilde {\Psi}}_{\theta,\varphi_0} \rangle= T_{Q,\varphi_0} R_{\theta} |\Psi_L \rangle$ with
\begin{equation}
R_{\theta}=e ^{ -\frac{\theta \sqrt{L}}{2}({\cal S}^+_0 - {\cal S}^-_0)},
\;\;\;\;\;
T_{Q,\varphi_0} = e ^{i \sum_j (jQ+\varphi_0) S^z_j}.
\end{equation}
$R_{\theta}$ is the global spin rotation by an angle $\theta$ along the $y$-axis, while 
$T_{Q,\varphi_0}$ is the ``twist'' operator with momentum $Q$ plus a global rotation by an 
angle $\varphi_0$ along the $z$-axis. In other words, $T_{Q,\varphi_0}$ rotates the spin $j$
by an angle $\varphi_j = Q j+\varphi_0$ along the $z$-axis. This implies that $|{\tilde {\Psi}}_{\theta,\varphi_0} \rangle$
is a canted spiral solution in wich the spin $j$ is fully polarized along the direction
${\bf u}_j = (-\sin{\theta} \cos{\varphi_j}, \sin{\theta} \sin{\varphi_j}, \cos{\theta})$ (see Fig.\ref{zz}b):
\begin{equation}
{\bf S}_j \cdot {\bf u}_j |{\tilde {\Psi}}_{\theta,\varphi_0} \rangle= \frac{1}{2}|{\tilde {\Psi}}_{\theta,\varphi_0} \rangle.
\label{pr}
\end{equation}
This implies that $|{\tilde {\Psi}}_{\theta,\varphi_0} \rangle$ is a direct product state:
\begin{eqnarray}
|{\tilde {\Psi}}_{\theta,\varphi_0} \rangle = \bigotimes_{j=0,L-1} |{\tilde {\psi}}_{\theta,\varphi_0} \rangle_j,
\label{dirp1}
\end{eqnarray}
where
\begin{eqnarray}
|{\tilde {\psi}}_{\theta,\varphi_0} \rangle_j = e^{i\frac{\varphi_j}{2}} \cos{\theta/2} |\!\uparrow \rangle_j 
+ e^{-i\frac{\varphi_j}{2}}\sin{\theta/2}
|\! \downarrow \rangle_j,
\label{dirp2}
\end{eqnarray}


Besides breaking the continuous U(1) symmetry of $H$, this solution breaks 
two discrete symmetries: spatial inversion, ${\cal I}$, and the product of time reversal times a $\pi$-rotation
along the $z$-axis ${\cal T} e^{i\pi \sqrt{L} {\cal S}_0^z}$. The more physical implication is that the canted spiral
carries non-zero vector, ${\boldsymbol \kappa}_j={\bf S}_j \times {\bf S}_{j+1} \cdot {\hat {\bf z}}$, and scalar,
${\boldsymbol \chi}_j= {\bf S}_{j-1} \cdot {\bf S}_j \times {\bf S}_{j+1}$, spin chiralities:
\begin{eqnarray}
\!\!\!\!\!\!\langle  {\tilde {\Psi}}_{\theta,\varphi_0} |{\boldsymbol \kappa}_j |{\tilde {\Psi}}_{\theta,\varphi_0} \rangle &=&
\frac{ 1}{4}\sin{Q} \sin^2{\theta},
\nonumber \\
\!\!\!\!\!\!\langle  {\tilde {\Psi}}_{\theta,\varphi_0} | {\boldsymbol \chi}_j
|{\tilde {\Psi}}_{\theta,\varphi_0} \rangle &=&
 \cos{\theta}\sin^2{\!\!\theta} \sin^3{\!\!(Q/2)} \cos{\!(Q/2)} .
\end{eqnarray}
These identities are easily obtained by exploiting the   .
direct product form of $|{\tilde {\Psi}}_{\theta,\varphi_0} \rangle$ [see Eq.(\ref{dirp1})]: 
$\langle  {\bf S}_j \times {\bf S}_l \rangle =\langle {\bf S}_j\rangle \times \langle {\bf S}_l\rangle$.
The coexistence of both chiralities implies that there are spin and electric currents circulating around each of the 
triangular plaquettes (see Fig.\ref{zz}a) \cite{Lev08,Khaled09}.

{\it Emergent SU(2) symmetry.} A spontaneously broken continuous symmetry is uncommon for $d=1$
systems because it is usually prohibited by the Mermin-Wagner theorem \cite{Mermin66}. The simplest counter-example
is the SU(2) ferromagnet ($Q=0$). In that case the continuous SU(2) symmetry  is spontaneously
broken  because the order parameter, $\sqrt{L} ({\cal S}^x_0, {\cal S}^y_0,{\cal S}^z_0)$ , 
coincides with the infinitesimal generators of the SU(2) symmetry group, i.e., it commutes with the Hamiltonian. 
Note that ${\cal S}^x_q = ({\cal S}^+_q+{\cal S}^-_q)/2$  and ${\cal S}^y_q = i({\cal S}^-_q-{\cal S}^+_q)/2$.
The situation is less clear for the canted spiral under consideration
($Q\neq0$) because the order parameter ${\bf P}_Q=(P^x_Q,P^y_Q,P^z_Q)=\sqrt{L} ({\cal S}^x_Q, {\cal S}^y_Q,{\cal S}^z_0)$ 
does not commute with $H$. 
However, we will show below that the components of the spiral order parameter generate an SU(2) group that is an 
{\it emergent} symmetry of $H$ \cite{Batista04}. 

The fact that the components of the canted spiral order parameter 
${\bf P}_Q=(P^x_Q,P^y_Q,P^z_Q)=\sqrt{L}({\cal S}^x_Q, {\cal S}^y_Q,{\cal S}^z_0)$ 
are elements of an su(2) algebra results from the commutation relations (\ref{cr}):
\begin{equation}
[P^\eta_Q,P^\mu_Q] = i \epsilon_{\eta \mu \nu} P^{\nu}_Q,
\label{cr2}
\end{equation}
where $\epsilon_{\eta \mu \nu} $ are the components of the Levi-Civita tensor.
According to Eq.(\ref{su2}), ${\bf P}_Q$ commutes with ${\cal P}_{{\cal G}_Q} H {\cal P}_{{\cal G}_Q}$  and this implies that the 
the SU(2) group generated by ${\bf P}_Q$ is an {\it emergent} symmetry of $H$ \cite{Batista04}. Moreover, the ground state subspace
${\cal G}_Q$ is an irreducible representation of this SU(2) group with an eigenvalue of the Casimir operator
$ {\bf P} \cdot {\bf P} |\Psi_p \rangle = L (L/4+1/2)  |\Psi_p \rangle$ $\forall 0 \leq p \leq L$. This eigenvalue coincides with the
square of the total spin for the uniform $Q=0$ case.

{\it Low Energy Excitations.} The U(1) and the translational invariance of $H$ imply that 
the single and two-magnon excitations on top of the fully polarized ground states, 
$|\emptyset \rangle$ and $|\Psi_L \rangle$, can also be computed in an
exact way. In particular, the wave function for the single magnon with wave-vector $q$
on top of the fully polarized state $|\emptyset \rangle$ (the other one is obtained by applying the time reversal symmetry) is
$
{\cal S}^+_q |\emptyset \rangle 
$,
and the corresponding energy eigenvalue is 
\begin{equation}
\omega_q = \sum_{\nu=1,2} J_{\nu} (\cos{\nu q}-\cos{\nu Q}). 
\end{equation}

{\it Two dimensions.} Eq.(\ref{klein}) shows that $H$ is a sum
of projectors $P_j$  over all the triangular plaquettes of the zigzag ladder. This structure suggest a natural extension of $H$ to
the $d=2$ case of parallel chains coupled by zigzag bonds (Fig.{\ref{zz}c}):
\begin{equation}
H^{2d} = \sum_{\bf r} H^{2d}_{{\bf r}^+} +H^{2d}_{{\bf r}^-} = (J_2+J_1^2/8J_2) \sum_{\bf r} (P_{{\bf r}^+} + P_{{\bf r}^-}),
\label{h2d}
\end{equation}
where
\begin{equation}
H^{2d}_{{\bf r}^{\pm}}= h_2 ({\bf r} \pm {\bf e}_1, {\bf r} \pm {\bf e}_2)
+ h_1 ({\bf r}\pm {\bf e}_{1}, {\bf r})
+ h_1 ({\bf r}, {\bf r} \pm {\bf e}_{2})
\end{equation}
and $P_{{\bf r}^{\pm}}$ is the projector on the subspace ${\cal E}({\bf r} \pm {\bf e}_1, {\bf r}, {\bf r} \pm {\bf e}_2)$ [see Eq.(\ref{subs})].
The sites ${\bf r}$ belong to the $L \times L/2$ lattice depicted in Fig.\ref{zz}c.
Note that $H^{2d}$ is an XXZ Hamiltonian with exchange interactions $J_1^z$, $J_1$ along the diagonal bonds and $2J_2^z$, $2J_2$ along 
the horizontal bonds (now each horizontal bond is common to two triangular plaquettes). Again, the boundary 
conditions are periodic. The three sites 
$({\bf r} \pm {\bf e}_1, {\bf r}, {\bf r} \pm {\bf e}_2)$ are the corners of the triangles that are above ($+$) and below
($-$) the site ${\bf r}$ (see Fig.\ref{zz}c).

The $d=2$ version of the canted spiral state is
\begin{equation}
|{\tilde {\Psi}}^{2d}_{\theta,\varphi_0} \rangle = \bigotimes_{\bf r} |{\tilde {\psi}}_{\theta,\varphi_0} \rangle_{\bf r},
\end{equation}
where $|{\tilde {\psi}}_{\theta,\varphi_0} \rangle_{\bf r}$ is given by Eq(\ref{dirp2}) with $\varphi_j$ replaced by 
$\varphi_{\bf r}= \varphi_0 + {\bf Q} \cdot {\bf r}$.
${\bf Q}$ is a wave-vector of the BZ that satisfies:
\begin{equation}
{\bf Q} \cdot {\bf e}_2 = - {\bf Q} \cdot {\bf e}_1 = Q.
\end{equation}
Again Eq.(\ref{cos}) holds for $Q=2\pi n/L$. 
To prove that $|{\tilde {\Psi}}^{2d}_{\theta,\varphi_0} \rangle$ is a ground
state of $H^{2d}$ we simply use the right hand side of Eq.(\ref{h2d}) and note that 
$|{\tilde {\psi}}_{\theta,\varphi_0} \rangle_{{\bf r}\pm {\bf e}_1} \otimes |{\tilde {\psi}}_{\theta,\varphi_0} \rangle_{\bf r} \otimes
|{\tilde {\psi}}_{\theta,\varphi_0} \rangle_{{\bf r} \pm {\bf e}_2}$  belongs to 
${\cal E}^{\perp}({\bf r} \pm {\bf e}_1, {\bf r}, {\bf r} \pm {\bf e}_2)$. This argument also provides an alternative way of proving that 
$|{\tilde {\Psi}}_{\theta,\varphi_0} \rangle$ is the ground state in the $d=1$ case of $H$.

{\it Conclusions.} In summary, we have demonstrated that the canted spiral solution is the ground state of 
a family of $S=1/2$ XXZ Hamiltonians defined on the zigzag ladder and parametrized by the wave-vector $Q$ of the spiral.
The spontaneous breaking of a continuous symmetry is very unusual for $d=1$ quantum Hamitlonians  when the order parameter
does not commute with $H$. We have shown that the case under consideration has an emergent SU(2) symmetry generated by the components
of the canted spiral order parameter.  

The family of exact canted spiral solutions includes the usual SU(2) ferromagnet in the $Q=0$ limit.
Like in the ferromagnetic case, the $Q \neq 0$ solutions are product states (unentangled), meaning that inter-site quantum fluctuations
that typically remove the long range spiral order \cite{Mermin66} are absent in this case. Moreover, the exact low-energy spectrum of
single magnon excitations exhibits a quadratic dispersion around the ordering wave-vectors $\pm Q$. 
In this sense, we can say that our canted spiral ground states are natural extensions of the ferromagnetic solution of 
isotropic Heisenberg models. Note also that our family of 
quasi-exactly solvable Hamiltonians includes cases that are arbitrarily close to the isotropic limit for $0< |Q| \ll \pi$ ($J_1 \lesssim -4 J_2$),
i.e., the exchange uniaxial anisotropy is a small perturbation. This regime corresponds to a ferromagnetic nearest-neighbor interaction $J_1<0$
($J_2$ is always antiferromagnetic) that has also been considered in previous 
works \cite{Somma01,Dimitriev08,Furukawa08,Heidrich-Meisner09}.

We have also shown how to extend our exact solutions to the two dimensional case of  XXZ Hamiltonians defined on a lattice that comprises
horizontal chains coupled by zig-zag bonds (see Fig.\ref{zz}c). This is just a spatially anisotropic triangular lattice 
that appears in real quantum magnets such as Cs$_2$CuCl$_4$ \cite{Coldea01}. By showing that $H$ can be written as a sum of projectors 
$P_j$ over triangular plaquettes [see Eq.(\ref{klein})], we are 
giving a prescription for constructing other higher dimensional Hamiltonians 
with exact canted spiral ground states.


This work was carried out under the  auspices of the NNSA
of the U.S. DOE at LANL under Contract No. DE-AC52-06NA25396.

\end{document}